\pdfoutput=1 
\documentclass[review]{mandm}

\usepackage{placeins}
\usepackage[utf8]{inputenc}
\usepackage{graphicx}
\usepackage{dcolumn}
\usepackage{bm}
\usepackage{hyperref}
\usepackage[version=4]{mhchem}
\newcommand{\ceil}{\mathrm{ceil}}

\usepackage{array}
\usepackage{algorithm,algcompatible}
\algnewcommand\algorithmicto{\textbf{to}}

\usepackage{hyperref}
\usepackage{adjustbox}
\usepackage{subcaption}
\usepackage{float}
\usepackage{placeins}
\usepackage{stfloats}

\usepackage{hyperref,color}
\definecolor{linkColor}{rgb}{1,0,0}
\hypersetup{pdfborder={0 0 0},colorlinks=true,urlcolor=linkColor,citecolor=linkColor}

\begin{document}
	
	\newcommand{\e}[1]{$\times 10^{#1}$}
	\newcommand{\vc}[1]{\mathbf{#1}}
	\newcommand{\Smat}{\mathcal{S}}
	\newcommand{\Amat}[0]{\mathcal{A}}
	\newcommand{\FFT}[4]{\mathcal{F}^{#4}_{\vc{#2}\rightarrow\vc{#3}}\left\{#1\right\}}
	\newcommand{\PIO}{PIO}
	\newcommand{\ZrO}{YSZ}
	\newcommand{\delete}[1]{}
	
	\title{A single-projection three-dimensional reconstruction algorithm for scanning transmission electron microscopy data}
	
	\author[Hamish G. Brown et al]{Hamish G. Brown,$^1$ \footnote{Current address: Ian Holmes Imaging Centre, Bio21 Molecular Science and Biotechnology Institute, the University of Melbourne, Parkville, Victoria 3052, Australia.}
		Philipp M. Pelz,$^{1,2}$ 
		Shang-Lin Hsu,$^{1,2}$
		Zimeng Zhang,$^2$
		Ramamoorthy Ramesh,$^{2,3}$
		Katherine Inzani,$^{4,5}$
		Evan Sheridan,$^{4,5,6}$
		Sin\'ead M. Griffin,$^{4,5}$
		Marcel Schloz,$^7$
		Thomas C. Pekin,$^7$
		Christoph T. Koch,$^7$
		Scott D. Findlay,$^8$
		Leslie J. Allen,$^9$
		Mary C. Scott,$^{1,2}$
		Colin Ophus,$^1$
		Jim Ciston$^1$
	}
	\affiliation{
		$^1$National Center for Electron Microscopy Facility, Molecular Foundry, Lawrence Berkeley National Laboratory, Berkeley, California 94720, USA\\
		$^2$Department of Materials Science and Engineering, University of California, Berkeley, CA, 94720, USA\\
		$^3$Department of Physics, University of California, Berkeley, California 94720, USA.\\
		$^4$Materials Sciences Division, Lawrence Berkeley National Laboratory, Berkeley, California 94720, USA\\
		$^5$Molecular Foundry, Lawrence Berkeley National Laboratory, Berkeley, California 94720, USA\\
		$^6$Theory and Simulation of Condensed Matter, Department of Physics, King's College London, The Strand, London WC2R 2LS, UK.\\
		$^7$Department of Physics \& IRIS Adlershof, Humboldt-Universit\"at zu Berlin, Newtonstra\ss e 15, 12489 Berlin, Germany
		$^8$School of Physics and Astronomy, Monash University, Victoria 3800, Australia\\
		$^9$School of Physics, University of Melbourne, Parkville, Victoria 3010, Australia
		Corresponding Authors: Hamish G Brown \email{hgbrown@unimelb.edu.au}, Colin Ophus \email{clophus@lbl.gov}, Jim Ciston \email{jciston@lbl.gov}}

	\date{\today}
	
	\begin{abstract}
		Increasing interest in three-dimensional nanostructures adds impetus to electron microscopy techniques capable of imaging at or below the nanoscale in three dimensions.
		We present a reconstruction algorithm that takes as input a focal series of four-dimensional scanning transmission electron microscopy (4D-STEM) data. 
		We apply the approach to a lead iridate, Pb$_2$Ir$_2$O$_7$, and yttrium-stabilized zirconia, Y$_{0.095}$Zr$_{0.905}$O$_2$, heterostructure from data acquired with the specimen in a single plan-view orientation, with the epitaxial layers stacked along the beam direction. We demonstrate that Pb-Ir atomic columns  are visible in the uppermost layers of the reconstructed volume.
		We compare this approach to the alternative techniques of depth sectioning using differential phase contrast scanning transmission electron microscopy (DPC-STEM) and multislice ptychographic reconstruction. 
		\noindent(Received XX Y 20ZZ; revised XX Y 20ZZ; accepted XX Y 20ZZ)
	\end{abstract}
	
	\maketitle
	\section{Introduction}
	There has recently been significant interest in nanoscale three-dimensional materials such as polarization vortices in layered \ce{PbTiO3}-\ce{SrTiO3} heterostructures~\citep{yadav2016observation}, van der Waals heterostructures~\citep{withers2015light} and strain-engineered nanoparticles~\citep{oh2020design}.
	Development of these materials is facilitated by imaging techniques capable of three-dimensional characterisation at nanometer resolution.
	However, three dimensional imaging techniques such as electron tomography~\citep{yang2017deciphering} and current optical sectioning methods in scanning transmission electron microscopy (STEM)~\citep{van2006three, xin2009aberration} typically rely on high-angle annular dark-field (HAADF) STEM which is often only sensitive to the heavy atoms within the sample.
	
	Techniques which reconstruct the electric potential of a sample from diffraction plane measurements of the modification of the electron probe after transmission through the specimen, such as differential phase contrast (DPC) STEM~\citep{shibata2017electric} and STEM ptychography~\citep{nellist1995resolution,yang2016simultaneous,jiang2018electron}, are alternatives that are sensitive to both light and heavy atoms. 
	The take-up of these techniques has been accelerated by recent advances in segmented detectors~\citep{shibata2010new} and the 4D-STEM technique -- where full two dimensional diffraction patterns are recorded for a two dimensional raster scan of a STEM probe. The latter has been enabled by advanced electron cameras capable of reading out full diffraction patterns at frequencies of the order of 100s of Hertz or greater, approaching the typical dwell times of a focused STEM probe, which typically vary from tens of $\mu$s to a few ms~\citep{ophus2019four}.
	
	In their most common implementations, ptychography and DPC assume that electron beam-sample interaction occurs in a single spatial plane -- referred to as the projection or phase object approximation -- a theoretical framework that is inconsistent with atomic resolution imaging of crystalline materials more than a few nanometer thick.
	Not withstanding the underlying projection approximation, DPC and phase object ptychography have both been usefully applied for three-dimensional imaging. 
	Since the final reconstruction is most sensitive to features close to where the beam comes to a focus, three dimensional information can be intuited from a DPC focal series of a specimen~\citep{bosch2019analysis}.
	Information about the three dimensional structure of a specimen has been demonstrated in single-sideband ptychography (SSB)~\citep{yang2016simultaneous} by varying in the reconstruction the $z$ coordinate of the single spatial plane at which beam-specimen interaction is assumed to have occurred.
	
	Multislice ptychography, an extension of the ptychography technique that will be used as a standard of comparison in this paper, has been proposed as a solution to the limitations of phase-object based ptychography. Here depth-wise propagation and multiple scattering of the illumination are taken into account by assuming that the illumination interacts with the sample at $n$ distinct depth (z) separated planes~\citep{maiden2012ptychographic}. 
	Though numerical stability of the algorithm remains problematic, requiring careful parameter selection and regularization~\citep{jiang2018breaking}, electron multislice ptychographic reconstruction has been demonstrated in experiment to improve in-plane resolution and reconstruction fidelity of thick materials~\citep{Schloz2020,chen2021electron} relative to phase-object based ptychography. 
	Optical sectioning of a carbon nanotube sample with nanotubes located at different depths~\citep{gao2017electron} has also been shown with the technique, demonstrating its promise as a three-dimensional reconstruction technique.

	This paper focuses on a three-dimensional imaging technique capable of visualizing both light and heavy atoms in a thick sample with nanoscale depth selectivity. 
	We reconstruct from a focal series of 4D-STEM measurements the scattering matrix ($\Smat$-matrix), a mathematical formalism common in quantum scattering theory~\citep{weinberg1995quantum} for calculating the phase and intensity of the scattered wave from a known input wave and that is capable of describing multiple electron scattering in an electron microscopy sample~\citep{sturkey1962calculation}. 
	We then use this $\Smat$-matrix to synthesize images of the specimen at different focal planes.	
	We apply this technique to a lead iridate, \ce{Pb2Ir2O7} (\PIO), and yttrium-stabilized zirconia, Y$_{0.095}$Zr$_{0.905}$O$_2$ (\ZrO), 
	heterostructure and demonstrate that Pb-Ir atomic columns are visible in the uppermost layers of the reconstructed volume.
	This builds on previous work~\citep{spence1998direct,allen2000inversion,findlay2005quantitative,brown2018structure,donatelli2020inversion} on $\Smat$-matrix retrieval that focused on single crystal structures.
	We compare results with a DPC-STEM focal series, phase-object ptychography and multislice ptychography. 
	For the case presented, only $\Smat$-matrix reconstruction led to identification of the Pb-Ir columns, which are identified by caldera-like features characteristic of heavy atoms in phase contrast reconstruction.
	%
	\section{Materials and methods}
	\subsection{Theory of $\Smat$-matrix reconstruction}
	For a brief derivation of the $\Smat$-matrix formalism we take as our starting point the Schr\"odinger equation in reciprocal space for the electron wave function with a relativistic velocity along the $z$-direction, for which the paraxial approximation is appropriate. 
	On a discretized grid with periodic boundary conditions~\footnote{This is exact for a crystal and a good approximation for a non-periodic object if the field of view is sufficiently large in real space.}, this equation is given by~\citep{coene1990inelastic}
	\begin{align}
	\frac{\partial \psi_\vc{g}}{\partial z} = -i \pi \lambda g^2\psi_\vc{g}(z) + \sum_h i \sigma V_{\vc{g}-\vc{h}}(z)\psi_\vc{h}(z)\,.
	\label{eq:shrodinger}
	\end{align}
	Here the Fourier coefficients of the electron wave function are given by $\psi_\vc{g}$ for reciprocal space coordinate $\vc{g}$ with amplitude $g$, $\lambda$ is the electron wavelength, $\sigma$ is the interaction constant\footnote{The interaction constant is given by $\sigma= 2\pi m_e e \lambda/h^2$ where $m_e$, $e$ and $h$ are the mass of an electron, the charge of an electron and Planck's constant respectively.} in radians/V and $V_{\vc{g}-\vc{h}}(z)$ are the 2D Fourier coefficients of the projected electrostatic potential at each depth z in the sample.
	From Eq.~(\ref{eq:shrodinger}) we can construct a matrix first-order differential equation,
	\begin{align}
	\frac{\partial \boldsymbol{\psi} (z)}{\partial z} = i \Amat (z)\boldsymbol{\psi} (z),
	\label{eq:ODE}
	\end{align}
	where $\boldsymbol{\psi}$ is a column vector containing the Fourier coefficients of the electron wave function. We write the entries of the matrix $\Amat$, the structure matrix, for an entry corresponding to the Fourier coefficient $\vc{g}$ of the scattered electron wave and Fourier coefficient $\vc{h}$ of the incoming electron wave function as,
	\begin{align}
	\Amat _{\vc{g},\vc{h}}(z)=-\pi \lambda g^2 \delta_{\vc{g}-\vc{h}}+\sigma V_{\vc{g}-\vc{h}}(z).
	\label{eq:structurematrix}
	\end{align}
	In the absence of any scattering potential (i.e. $V_{\vc{g}-\vc{h}}(z)=0$) the diagonal terms, $-\pi \lambda g^2 \delta_{\vc{g}-\vc{h}}$, where $\delta_{\vc{g}-\vc{h}}$ is the Kronecker delta, have an equivalent effect to the Fresnel free-space propagator. 
	If, on the other hand, the terms $-\pi \lambda g^2$ are ignored then Eq.~(\ref{eq:structurematrix}) can be shown to reduce to the phase object approximation.
	Taking both sets of terms together, Eq.~(\ref{eq:structurematrix}) encapsulates the simultaneous probe-specimen interaction (via $V$) and propagation that leads to the phenomenon colloquially referred to as ``dynamical diffraction''.
	
	A standard solution to Eq.~(\ref{eq:structurematrix}) is the $\Smat$-matrix solution,
	\begin{align}
	\boldsymbol{\psi}(z+\Delta z)=e^{i\Delta z\Amat (z)}\boldsymbol{\Psi}(z) = \Smat(\Delta z)\boldsymbol{\Psi}(z)\,.
	\label{eq:scatteringmatrix}
	\end{align}
	It is implicitly assumed in Eq.~(\ref{eq:scatteringmatrix}) that $ V_{\vc{g}-\vc{h}}(z)$ is constant over thickness $\Delta z$. 
	Where $ V_{\vc{g}-\vc{h}}(z)$ varies with thickness, the $\Smat$-matrix can be constructed as a product of $\Smat$-matrices over $n$ thinner sub-regions within which the $z$ variation of $ V_{\vc{g}-\vc{h}}(z)$ is minimal,
	\begin{align}\label{eq:Nslices}
	\Smat(z)=\prod_{i=0}^{n-1} \Smat(z_{i+1}-z_i).
	\end{align}
	In a 4D-STEM experiment the diffraction pattern for each raster scan position of a focused probe is measured and, for a probe position $\mathbf{R}$ and diffraction coordinate $\vc{g}$, this diffraction pattern can be calculated using the $\Smat$-matrix as 
	\begin{align}
	I(\vc{g},\vc{R},\Delta f) =\left |\sum_{\vc{h}}\Smat_{\vc{g},\vc{h}}A(\vc{h})e^{-2\pi i \vc{h}\cdot \vc{R}-i \pi \lambda h^2\Delta f}\right|^2 \,.
	\end{align}
	Here $\Delta f$ is the probe defocus relative to the entrance surface of the specimen and $A(\vc{h})$ is the aperture function, a top-hat function that is 1 for Fourier components within the aperture, $|\vc{h}|<h_{\text{max}}=\alpha/\lambda$ ($\alpha$ is the probe convergence semi-angle in radians), and 0 otherwise. 
	We seek to recover the complex-valued $\Smat$-matrix for a set of real-valued 4D-STEM datasets $I(\vc{g},\vc{R},\Delta f)$ recorded for a number of scan positions $\vc{R}$ and defoci $\Delta f$, a phase retrieval problem which we solve using a gradient descent approach~\citep{Guizar-Sicairos_Fienup_2008,Thibault_Guizar-Sicairos_2012,wang2017solving,pelz2020reconstructing}, see Algorithm \ref{Alg:1}.
	\begin{algorithm*}
		\caption{$\Smat$-matrix retrieval from a 4D-STEM dataset via gradient descent using an amplitude difference cost function~\cite{wang2017solving,pelz2020reconstructing}\label{Alg:1}}
		Input: \\
		4D-STEM datasets $I(\vc{g},\vc{R},\Delta f)$, with pixel dimensions $(g_x,g_y,R_x,R_y,n_{\Delta f})$. We denote the total number of scan positions across all defoci $N_{\mathbf{R}_i,\Delta f}$\\
		Lens defoci $\Delta f$ in units of length\\
		Probe positions $\vc{R}$  in units of length\\
		Probe forming aperture $g_\alpha$ in units of inverse length\\
		$\mu$, the algorithm ``step size''\\
		\textbf{Initialize :} \\
		Calculate reconstruction grid dimensions $(M_x/\Delta g_x,M_y/\Delta g_y)$, where $M_i=\ceil{(\max(R_i)-\min(R_i))\Delta g_i}$ for $i=x,y$, $\Delta g_i$ is the diffraction pixel size in units of inverse length and $\ceil$ is the ceiling function.\\
		Initialize $\Smat$-matrix : $\Smat_{\vc{g},\vc{h}} = \delta_{\vc{g}-\vc{h}}$, where the the input Fourier coefficients are those that sit within the probe forming aperture, $\{\vc{h}\, :\, |\vc{h}|<g_\alpha \}$.\\
		Calculate illumination matrix for each probe position and defocus, $\vc{\psi}_{\vc{h},\{\Delta f,\vc{R}\}}(0) = Ae^{2\pi i \vc{R}\cdot \vc{h} -i\pi h^2 \lambda\Delta f}$, $A$ is the mean diffraction pattern amplitude in the 4D-STEM dataset $A = \sqrt{\sum_{\Delta f,\vc{R}}{I(\vc{g},\vc{R},\Delta f)}/n_{\Delta f}/R_x/R_y}$\\
		\textbf{Run:}
		\begin{algorithmic}
			
			\FOR{$\mathsf l=0$ to $\mathrm{L}$} \COMMENT{Loop over amplitude flow iterations}
			\FOR{$\mathbf{R}_i$, $\Delta f_i$ in \{$\mathbf{R}$, $\Delta f$\} } \COMMENT{Loop over scan postions and defoci in dataset}
			\STATE{
				$\hat{\Psi}(\vc{g},\vc{R}_i,\Delta f_i) = \sum_\vc{h}\Smat_{\vc{g},\vc{h}}\vc{\psi}_{\vc{h},\{\vc{R}_i, \Delta f_i\}}$} \COMMENT{Forward operation}
			\STATE{$\hat{\Psi}(\vc{g},\vc{R}_i,\Delta f_i) =\hat{\Psi}(\vc{g},\vc{R}_i,\Delta f_i)- I(\vc{g},\vc{R}_i,\Delta f_i)\cdot\hat{\Psi}(\vc{g},\vc{R}_i,\Delta f_i)/|\hat{\Psi}(\vc{g},\vc{R}_i,\Delta f_i)|$}
			\STATE{$\Smat_{\vc{g},\vc{h}} -=  \mu/N_{\Delta f,\mathbf{R}_i} \hat{\Psi}(\vc{g},\vc{R}_i,\Delta f_i)\otimes_{\text{outer}} \vc{\psi}^*_{\vc{h},\vc{R}_i,\Delta f_i}$} \COMMENT{Back-projection (adjoint) operation}
			\ENDFOR
			\ENDFOR
		\end{algorithmic}
	\end{algorithm*}
	%
	%
	%
	\subsection{Depth sectioning from the $\Smat$-matrix}

	The specimen potential $V(x,y,z)$ would ideally be directly retrieved from the $\Smat$-matrix resulting from the phase-retrieval step, and previous work has identified a quantitative method of doing this for a perfect crystalline sample~\citep{spence1998direct,allen2000inversion,donatelli2020inversion} and demonstrated the technique experimentally~\citep{brown2018structure}. 
	Here we consider the general case of a more heterogeneous sample such as a heterostructure or nanoparticle and we use an optical sectioning approach that estimates the potential at a given depth of the object \citep{ophus2019advanced}.
	
\begin{figure}
		\includegraphics{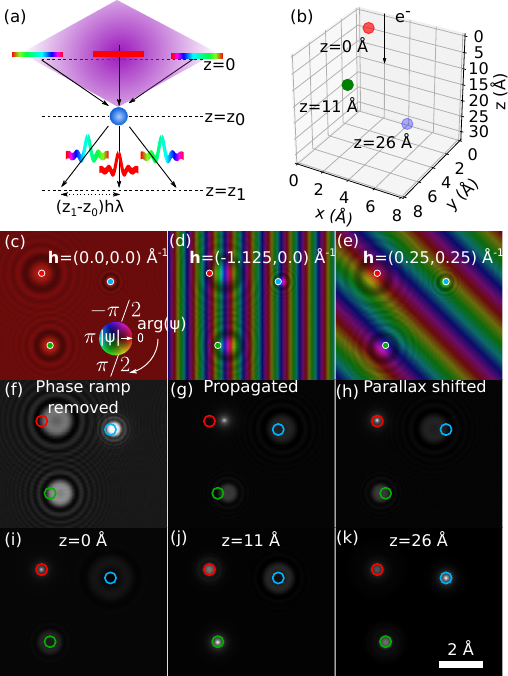}
		\caption{Reconstructing three dimensional information from the $\Smat$-matrix. (a) For an atom at depth $z_0$, each component of the $\Smat$-matrix will acquire a phase through interaction with this atom. For a toy model system of three oxygen atoms at different points in 3D space as shown in (b) the complex $\Smat$-matrix components for three different Fourier component inputs $\vc{h}$  are shown in (c-e) plotted with phase and intensity given by the color-wheel in (c). The reconstruction method described in the text involves (f) removing the phase ramp, (g) propagating the complex wave function back to the plane of interest and (h) correcting the parallax shift from propagation. Only the phase is plotted in gray-scale in these and following panels. Shown in (i)-(k) are the reconstructions for the respective planes of the different atoms. \label{fig:schematic}}
	\end{figure}

	Consider $\Smat_{\vc{r},\vc{h}}$ (the $\Smat$-matrix component that maps plane wave input $\vc{h}$ to a real-space exit surface wave function~\footnote{The $\Smat$-matrix is a mathematical construct common in quantum physics that maps the input momentum states (i.e. Fourier components of the wavefunction) of the incoming particle to the output momentum states~\citep{weinberg1995quantum}. In this paper we often use a mixed momentum and real space representation of this quantity which is related to the more conventional representation of the $\Smat$-matrix via an inverse Fourier transform operation ($\Smat_{\vc{r},\vc{h}}=\mathcal{F}^{-1}_{\vc{g}\rightarrow\vc{r}}\Smat_{\vc{g},\vc{h}}$).}) for the case indicated in Fig.\ref{fig:schematic}(a), freespace propagation through distance $z_0$, phase object interaction with potential (e.g. an atom), and further propagation through distance $\Delta z=z_1-z_0$,
	%
	%
	\begin{align}
	\Smat_{\vc{r},\vc{h}}&=\mathcal{P}(\vc{r},\Delta z)\otimes_{\vc{r}} \left[e^{i\sigma V(\vc{r})}e^{-i\pi\lambda h^2z_0}e^{2\pi i \vc{h}\cdot \vc{r}}\right]\,.
	\label{Eq:Smatcomponent}
	\end{align}
	Here $\mathcal{P}(\vc{r},z)$ is the real space representation of the Fresnel free space propagator for propagation of distance $z$ and the potential for the atom $V(\mathbf{r})$ is assumed projected (ie. depth integrated). 
	Using the convolution theorem, Eq.~(\ref{Eq:Smatcomponent}) can be written as the inverse Fourier transform of the product of the Fourier transforms of the individual expressions:
	\begin{align}
	\Smat_{\vc{r},\vc{h}}&=e^{-i\pi \lambda z_0h^2}\int{e^{-i\pi \lambda \Delta zg^2}\hat{T}_{\vc{g}-\vc{h}}e^{2\pi i \vc{g}\cdot \vc{r}}}d\vc{g}\,,
	\label{Eq:SmatFFT}
	\end{align}
	where we have defined $\hat{T}_\vc{g} = \FFT{e^{i\sigma V(\vc{r})}}{r}{g}{}$ and made use of the Fourier shift theorem. Making the change of variable $\vc{g}\rightarrow\vc{g}+\vc{h}$, Eq.~(\ref{Eq:SmatFFT}) becomes
	\begin{align}
	\Smat_{\vc{r},\vc{h}}=&\ e^{-i\pi \lambda z_0h^2}\int{e^{-i\pi \lambda \Delta z(\vc{g}+\vc{h})^2}\hat{T}_{\vc{g}}e^{2\pi i (\vc{g}+\vc{h})\cdot \vc{r}}}d\vc{g}\,,\\ 
	=&\ e^{2\pi i \vc{h}\cdot \vc{r}-i\pi \lambda z_1h^2}\int{e^{-i\pi \lambda \Delta zg^2}\hat{T}_{\vc{g}}e^{-2\pi i \vc{g}\cdot (\lambda\Delta z\vc{h}+\vc{g}\cdot\vc{r})}}d\vc{g}\,.
	\end{align}
	Invoking the convolution theorem and Fourier shift theorem again (in reverse form to that used previously), this becomes,
	\begin{align}
	\Smat_{\vc{r},\vc{h}}&=e^{2\pi i \vc{h}\cdot \vc{r}-i\pi \lambda z_1h^2}\left[\mathcal{P}(\vc{r},\Delta z)\otimes_{\vc{r}}e^{i\sigma V(\vc{r}-\vc{h}\lambda\Delta z)}\right]\,.
	\end{align}
	For this case, the $\Smat$-matrix consists of the atom with a lateral shift of $\vc{h}\lambda \Delta z$ and  free-space propagation of $\Delta z$ [schematically shown in Fig.~\ref{fig:schematic}(a)] with a multiplicative phase ramp $e^{2\pi i \vc{h}\cdot\vc{r}}$. 
	To generate our optical section reconstruction at depth $z$, we apply the inverse of each of these processes (i.e. the phase ramp, propagation and paraxial shift) for each $\vc{h}$ in $\Smat_{\vc{r,h}}$, sum $\Smat_{\vc{r},\vc{h}}$ over all $\vc{h}$ and the phase of the result should be a reasonable approximation to $V(\vc{r})$.
	Averaging over the different momentum components $\vc{h}$ of the STEM probe provides some robustness against the effects of multiple scattering, although it is not addressed explicitly in the depth sectioning part of the algorithm.
	This is analagous to the diminution of dynamical effects observed as a result of averaging diffraction patterns over different beam tilts in the precession electron diffraction technique~\citep{ciston2008quantitative}.
	
	Although the algorithm will be applied to a strongly scattering crystalline sample later in the paper, for clarity of explanation we will first demonstrate the algorithm on a simulated $\Smat$-matrix of a toy model consisting of three weakly scattering oxygen atoms at different depths as shown in Fig.~\ref{fig:schematic}(b).
	\delete{We demonstrate this approach on a toy model consisting of three oxygen atoms which is shown in Fig.~\ref{fig:schematic}(b).}
	Forward simulated $\Smat$-matrix components for different Fourier components $\vc{h}$ are shown in Fig.~\ref{fig:schematic}(c-e), plotted with phase and intensity given by the color-wheel in (c).
	The positions of the atoms projected onto the 2D plane are indicated with colored dots and the paraxial shift from the projected positions of each of these atoms is visible.
	Shown in the next row of Fig.~\ref{fig:schematic}, with now just the phase plotted in gray color scale,  are first the removal of the phase ramp $e^{2\pi i \vc{h}\cdot\vc{r}}$ in (f), the application of the propagation operator $\mathcal{P}(\vc{r},-32$ \AA $)$ in (g) and finally, in (h), the correction of the paraxial shift $\lambda (z_1-z_0) \vc{h}$ to reconstruct the atom at nominal height $z=0$ from Fig.~\ref{fig:schematic}(b).
	These steps are applied to all components of $\Smat_{\vc{r},\vc{h}}$ and summed over all $\vc{h}$, which as can be seen in Fig.~\ref{fig:schematic}(i) further diminishes the contributions of atoms at different depths to the plane of reconstruction.
	The process is repeated for the other two atoms in Fig.~\ref{fig:schematic}(j) and (k).
	The high spatial resolution exhibited in these reconstructions is the result of their reconstruction from a forward simulated $\Smat$-matrix -- finite signal-to-noise and partial coherence of the STEM probe will limit the fidelity of reconstruction in the experimental case.
	
	%
	\subsection{Multislice ptychography}
	In the formalism of ptychography, the phase object approximation is usually applied to the Schrödinger equation and the probe-specimen interaction is sufficiently described for thin specimens in the absence of dynamical scattering. 
	A single 2D slice $V(r_{x,y})$, onto which the specimen potential is projected along the z-axis, can then be reconstructed by the ptychographic reconstruction algorithm of choice. 
	For thick specimens that involve multiple scattering the projection approximation breaks down and reconstruction quality deteriorates. 
	Multislice ptychography aims to circumvent these issues. 
	Probe-specimen interaction is modelled by the multislice algorithm, a split step solution to the Schr\"odinger equation for a fast electron, Eq.~(\ref{eq:shrodinger}), where the specimen potential is projected into $N$ distinct slices, as in Eq.~(\ref{eq:Nslices}). Propagation and transmission operations are applied sequentially within those slices and an iteration of the multislice algorithm to advance the real space wave function $\psi_n(\vc{r_{x,y}})$ from slice $n$ to slice $n+1$ may be written:
	\begin{equation}
	\psi_{n+1}(\vc{r_{x,y}})=\mathcal{P}(\vc{r_{x,y}},\Delta z)\otimes_{\vc{r_{x,y}}} \left(\psi_n(\vc{r_{x,y}}) e^{i\sigma V_n(\vc{r_{x,y}})}\right)\,.
	\label{Eq:MSinteraction	}
	\end{equation}
	From the various existing ptychographic reconstruction algorithms that use the multislice method to recover the potential of a thicker specimen, we consider a gradient based algorithm with a Polak-Ribi\`ere conjugate gradient optimization method. Here, a loss function is formed and iteratively minimized in the search direction that is guided by its gradients. The partial derivatives, required to generate the gradients, are efficiently calculated by backpropagation of the loss function. A more detailed description of the ptychographic reconstruction algorithm is given in~\cite{Schloz2020}.
	\subsection{4D-STEM dataset acquisition}
	\begin{figure}
		\centering
		\includegraphics[width=\columnwidth]{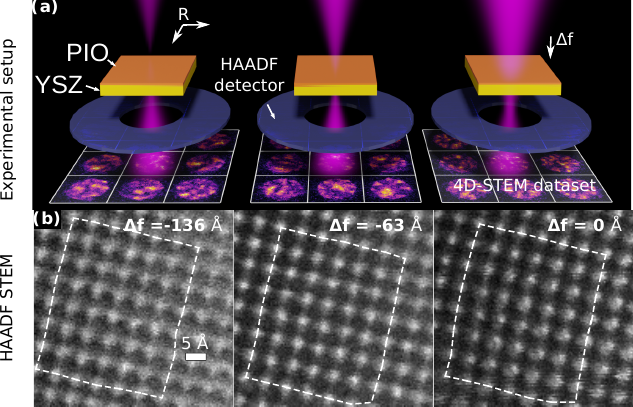}
		\caption{(a) Recording of a 4D-STEM focal series for $\Smat$-matrix reconstruction. Simultaneously acquired HAADF STEM images are shown in (b).
			\label{fig:HAADFdiagrm}}
	\end{figure}
	
	For an experimental demonstration, we used a \PIO\ layer grown on a (001)-oriented \ZrO\ substrate by pulsed laser deposition at a growth temperature of 600 $^\circ$C, wedge polished on the \ZrO\ substrate side and then  ion milled with a low-energy Ar$^+$ beam. 
	Experiments were performed using the TEAM I instrument at the National Center for Electron Microscopy (NCEM) Facility of the Molecular Foundry, a double aberration-corrected Thermo-Fisher Titan 80-300. 
	The instrument was operated with an accelerating voltage of 300 kV and a 20 mrad probe forming aperture. 
	The reconstructed $\Smat$-matrix is sensitive to residual probe aberrations so careful initial alignment of the probe corrector and constant tuning of the stigmators before acquisition of 4D-STEM focal series was necessary to achieve good results. 
	As shown in Fig.~\ref{fig:HAADFdiagrm}(a), a focal series of 4D-STEM data was recorded on a Gatan K3  direct-electron detector, operated in counting mode, at the end of a Gatan Continuum imaging filter with an energy slit width of 15 eV centered around the zero-loss peak.
	A probe step of 0.21 \AA\ with probe dwell time of 0.874 ms and beam current of 2.01 pA (estimated using the K3 camera) was used. 
	To maximise source coherence and minimize beam damage, source magnification (the ``spot size'' software setting) was set to 10, the penultimate setting. 
	HAADF STEM images, from a detector inner angle of 110 mrad, recorded concurrently with the 4D-STEM data are shown in Fig.~\ref{fig:HAADFdiagrm}(a).
	Alignment of the frames in the focal series was achieved by fitting two-dimensional Gaussian functions to the atomic columns in the STEM HAADF images using the open-source Atomap package~\citep{nord2017atomap} and smoothly deforming the probe positions to match a rectangular crystalline lattice rotated to the same average orientation as the fitted lattice. 
	This is detailed in Fig.~S1 of the supplementary material, the region that was input into the $\Smat$-matrix reconstruction is shown with a dashed white outline in the HAADF STEM results in Fig.~\ref{fig:HAADFdiagrm}(b). 
	Defocus values reported by the microscope software had to be adjusted by a multiplicative factor of 1.24, a value determined by comparing the geometric blur  with defocus of gold nanoparticles embedded in amorphous carbon with the geometric blur expected from a 20 mrad probe forming aperture (see  Fig.~S2 of the supplementary material).

	\section{Results}
	\subsection*{$\Smat$-matrix depth sectioning results}
	\begin{figure}[t]
		\includegraphics[width=\columnwidth]{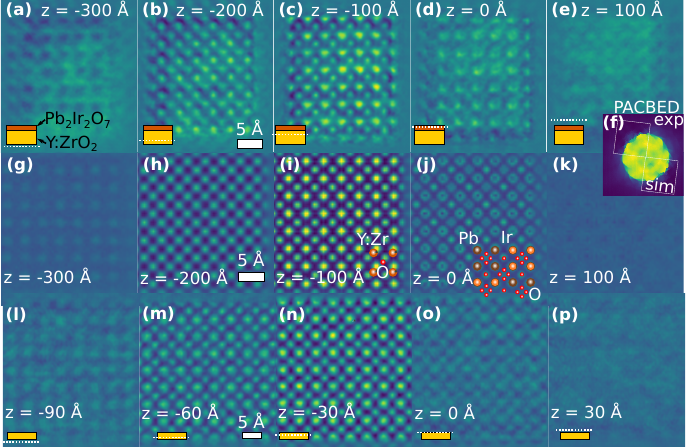}
		\caption{Experimental reconstruction of the $\Smat$-matrix from a focal series of 4D-STEM scans of a \PIO-\ZrO\ heterostructure.  The optical section reconstructed from experiment in Fig.~\ref{fig:HAADFdiagrm} is shown in (a)-(e) with the ~150 \AA\ \ZrO\ substrate visible in (b) and (c) and the ~50 \AA\ thick \PIO\ layer visible in panel (d). An experimental and simulated position-averaged convergent beam electron diffraction (PACBED) pattern  in (f) provides supporting evidence for the composition and structure implied by depth sectioning. Panels (g)-(k) are from a reconstruction from simulated data of such a structure. For reference, a similar $\Smat$-matrix depth section reconstructed from a 4D-STEM experiment with a ZrO only structure is shown in (l)-(p) and no ``caldera'' features to indicate Pb-Ir columns are evident.
			\label{fig:exp}}
	\end{figure}
	The experimental $\Smat$-matrix was reconstructed using 10 iterations of the gradient descent algorithm described in Algorithm~\ref{Alg:1}.
	Convergence with increasing iterations of the algorithm is shown in Fig.~S3.
	Select components of the $\Smat$-matrix are shown in Fig.~S4(a)-(e), and optical sectioning applied to the results.
	Optical sections at 100 \AA\ intervals are shown in Fig.~\ref{fig:exp}(a)-(e). The -300 \AA\ and 100 \AA\ sections are outside the bounds of the object, though the lattice is still faintly visible. The presence of Pb and Ir in the upper layers of the heterostructure is evidenced by appearance of ``caldera'' or volcano like atoms in Fig.~\ref{fig:exp}(d).
	It is commonly seen in phase reconstructions in STEM that the phase imparted to an electron wave by high Z atomic columns, which strongly elastically and inelastically scatter the electron probe, are observed to underestimate the true scattering potential of an object leading to a dip in the reconstructed phase close to the atomic position~\citep{close2015towards,yang2017electron}.
	Thus the Pb and Ir atoms (Z=82 and Z=77) are observed to be darker in the reconstruction than the lighter Y and Zr atoms (Z=39 and Z=40).
	From the optical section we estimate the thickness to be 200 \AA\ with a composition of approximately 50 \AA\ of \PIO\ and 150 \AA\ of \ZrO. This is supported by comparison of the scan position-averaged convergent beam electron diffraction (PACBED) pattern~\citep{lebeau2010position} from experiment with that simulated for a model structure of \PIO-\ZrO\ of that thickness and composition of both materials in (f).
	
	Fig.~\ref{fig:exp}(g)-(k) shows the results of simulating and then reconstructing a model structure with 200 \AA\ total thickness and containing 50 \AA\ of \PIO, and 150 \AA\ of \ZrO\ for equivalent focal conditions to the experimental data in Fig.~\ref{fig:exp}(a)-(e), showing good overall qualitative agreement with experiment.
	Inclusion of defocus spread due to chromatic aberration (assuming a full-width at half-maximum probe energy spread of 0.8 eV, and lens Cc coefficient of 1.4 mm for the TEAM 1 instrument) was the most important experimental effect necessary to include in simulation for good agreement. 
	As a demonstration how the appearance of caldera-like atoms is a signature of heavier atoms for this particular system, an experimental reconstruction from a different and much thinner region of the sample, free of \PIO, is shown in Fig.~\ref{fig:exp}(l)-(p). 
	Caldera-like atoms are not visible at any of the slices of the reconstruction, nor in any of slices of a reconstruction from equivalent simulated dataset (not shown).
	
	\begin{figure}
		\includegraphics[width=\columnwidth]{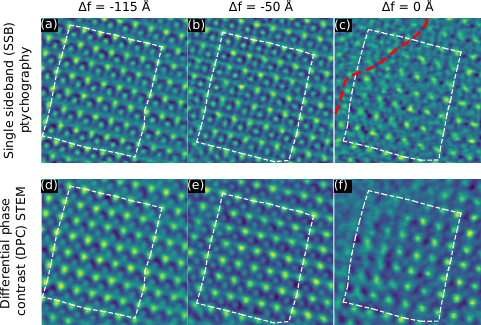}
		\caption{Phase object approximation reconstructions of the experimental data from (a)-(c) single side band ptychography (SSB)~\citep{rodenburg1993experimental,pennycook2015efficient,yang2016simultaneous} and (d)-(f) differential phase contrast (DPC) STEM~\citep{close2015towards,shibata2017electric}. These approaches generate broadly similar results and the \PIO\ layer is visible in the right-most panel ($\Delta f=0$ \AA) (c) for SSB, where the approximate bounds of the \PIO\ layer are indicated by a red dashed line, and (f) for DPC.\label{fig:SSBDPC}}
	\end{figure}
	\subsection*{Reconstructions from phase-object based approaches}
	Phase reconstructions from single sideband (SSB) ptychography STEM, calculated from the 4D-STEM dataset, are shown in ~\ref{fig:SSBDPC}(a)-(c). As was the case with the $\Smat$-matrix reconstruction, SSB reveals the oxygen columns in the \PIO -\ZrO\ structure which are invisible in the HAADF STEM images of Fig.~\ref{fig:HAADFdiagrm}(b) and suggests, in the $\Delta f = 0 $ \AA\ image, that the dataset was recorded in a region of the specimen where the \PIO\ layer terminates, with the upper left region of the image apparently only containing the \ZrO\ substrate.
	The approximate boundary between these regions is indicated by a red dashed line in Fig. 4(c).
	This is consistent with the optical section in Fig.~\ref{fig:exp}(d) showing stronger evidence of Pb and Ir atoms in the bottom right than in the upper left of the reconstruction.
	A differential phase contrast (DPC) STEM reconstruction of the datacube produces similar results to SSB ptychography, see Fig.~\ref{fig:SSBDPC}(d)-(f).
	%
	\begin{figure}
		\includegraphics[width=\columnwidth]{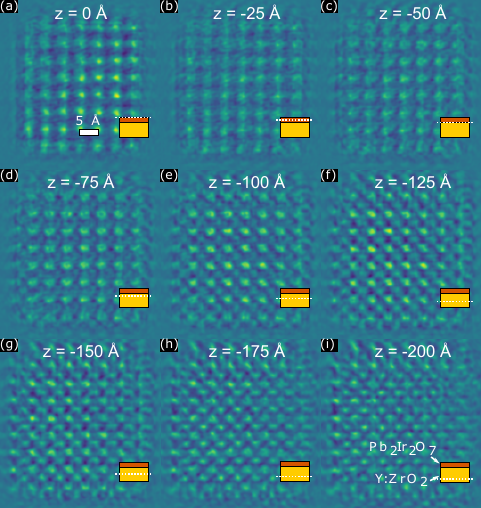}
		\caption{Multislice ptychography reconstruction from the \PIO - \ZrO\ experimental dataset with defocus $\Delta f=0$ \AA\ with respect to the specimen surface. The termination of the \PIO\ layer is evident in (a) whilst slices reconstructed from deeper within the crystal (b)-(j) suggest a uniform crystal lattice. \label{fig:msptycho}}
	\end{figure}
	\subsection*{Multislice ptychographic reconstruction}
	The electrostatic potential slices obtained by the multislice ptychographic reconstruction are shown in Fig.~\ref{fig:msptycho} for the 4D-STEM experimental dataset recorded simultaneous with the $\Delta f=0$ \AA\ HAADF-STEM image from Fig.~\ref{fig:HAADFdiagrm}(b). 
	Optimization was done for 400 iterations, alternating between 2 iterations of potential update and 6 iterations of probe shape update, respectively. For the loss function, the $\ell_1$ error metric was employed and no regularization has been used.
	The potential was reconstructed in 10 distinct planes seperated from one another by 25 \AA\ in the $z$ direction which are all displayed in Fig.~\ref{fig:PIOdisorder}(a)-(j). 
	Results are broadly in agreement with the conclusions drawn from the $\Smat$-matrix, SSB and DPC-STEM analyses. 
	The termination of the \PIO\ flake is visible in the uppermost-slice of the reconstruction Fig.~\ref{fig:msptycho}(a) and panels (b)-(i) are suggestive of a more uniform crystal lattice, showing that a single defocus reconstruction gives accurate 3D information about the crystal. 
	Three seperate defoci were required to achieve comparable insights in Fig.~\ref{fig:SSBDPC} using SSB and DPC-STEM analyses.
	
	The appearance of the cation columns is noticeably different within the \PIO\ layers, Fig.~\ref{fig:msptycho}(b) and (c), though this insight only seems well supported if the results of the other techniques employed within this paper are taken into account. 
	Fig.~\ref{fig:msptycho} (h) and (i) also look noticeably distinct from the other planes, though we suspect that this is an artefact of reconstructing planes located far from the focal point of the STEM probe. 
	The scattering of the probe is more strongly determined by atoms close to where the probe comes into focus than by planes where the probe has spread significantly.
	This becomes clearer when viewed alongside reconstructions from the other datasets in Fig.~S5 of the supplementary material, reconstructed planes immediately below the focal plane in the multislice ptychography reconstruction are the sharpest and those far above or below the focal plane ($\pm$ 150 \AA) are significantly blurrier.
	To overcome this issue, multislice ptychography reconstruction algorithms should be designed to take into account datasets with multiple defoci as the $\Smat$ algorithm does and we highlight this as an area for future work.
	Though all the reconstruction techniques applied to the 4D-STEM dataset are supportive of similar conclusions, at present only the $\Smat$-matrix reconstruction gives definitive evidence of a \PIO\ layer in the upper-layers of the volume.
	\subsection*{Comments on the appearance of oxygen columns in the PIO structure}
\begin{figure}
		\includegraphics[width=\columnwidth]{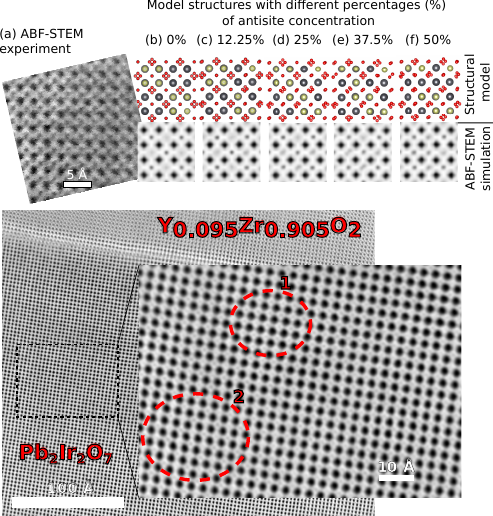}
		\caption{The annular bright field (ABF) STEM image synthesised from the 4D-STEM dataset of Fig. 2 ($\Delta f =0$ \AA) lacks the alternating ``splayed'' and ``tight'' ordering of the oxygen columns that is predicted in the \PIO\ pyrochlore structure in sub-figure (b). Increasing the amounts of cation disorder, (c) 12.25\%, (d) 25\%, (e) 37.5\% and (f) 50\% antisites per unit cell, disrupt the oxygen columns somewhat, though even with very high levels of disorder a satisfactory match with experiment is not achieved. Shown in (b) are ABF STEM images of a sample prepared in cross-section. In the zoomed view regions of alternating localized and delocalized oxygen columns are seen (e.g. region 2) as well as regions where the alternating localization and delocalization of the columns is not visible. \label{fig:PIOdisorder} }
	\end{figure}

	A final point to note is that PIO is a pyrochlore structure and, when projected down its [001] crystallographic zone-axis, the oxygen columns should exhibit alternating centered and delocalized columns, as shown in the structural overlay of Fig.~\ref{fig:exp}(j). 
	This was not visible in this reconstruction of the Pb$_2$Ir$_2$O$_7$ structure.
	The alternating delocalized O columns were also not visible in any of the annular bright-field (ABF) STEM images synthesized from the 4D-STEM datasets, e.g. see Fig.~\ref{fig:PIOdisorder}(a) for the dataset displayed in Fig.~\ref{fig:exp}, which rules out the effect being an artefact of any of the reconstruction algorithms and makes it clear that this is a real feature of the as-prepared sample that requires some explanation.
	
	To account for this, density functional theory (DFT) calculations were performed to determine the extent of compositional and structural influences on the oxygen positions.
	We use the parameter $x$ to describe the position of O on the Wyckoff site 48$f$ ($x$,$\frac{1}{8}$,$\frac{1}{8}$) of the Fd$\overline{3}$m (No. 227-2) space group, where $x$=0.375 gives a centered oxygen column and $x$=0.3125 is the maximum amount of oxygen off-centering allowed in the pyrochlore structure~\citep{Subramanian1983}. Values of $x$ for different structural variations are given in Table~\ref{tab:dft} (details in the supplementary material). Incorporating biaxial strain, stoichiometry variations (through the inclusion of oxygen vacancies) and an explicit YSZ/PIO interface do not cause significant changes in $x$. For 37.5\% cation antisite defects, there is a considerable increase in $x$, indicating that a large proportion of antisite defects could induce a noticable change to the alternating pattern. Simulated STEM images of this amount of disorder are shown in Fig.~\ref{fig:PIOdisorder}(b)-(f). 
	We see that some individual oxygen columns are contracted.
	However, such a high proportion (37.5\% and above) of cation antisite defects seem unlikely - we hypothesise that some smaller amount of antisite disorder is likely a factor contributing to the lack of localized-delocalized oxygen column ordering but cannot, on its own, fully explain the observed phenomenon.
	ABF-STEM images from a sample of the same material prepared such that the \PIO-\ZrO\ interface could be viewed in cross-section revealed regions where the alternating centered and delocalized oxygen columns were visible and areas where adjacent oxygen columns appeared to be centered, see Fig.~\ref{fig:PIOdisorder}(g). 
	This suggests this alternating oxygen column order is likely present in nano-domains within the sample and cation disorder, either from the growth process or damage during imaging, results in alternating order and is not noticeable when the sample is viewed in plan-view projection in the electron microscope.
	\begin{table}[ht]
		\begin{tabular}{lc}
			\hline
			\multicolumn{1}{l}{Structural variation }                          & \textit{x}    \\ \hline
			\ce{Pb2Ir2O7}                                                  & 0.330         \\
			\ce{Pb2Ir2O7} biaxially strained to 10.28 \AA                  & 0.332             \\
			PIO(001)/YSZ(001) interface                               & 0.332             \\
			\ce{Pb2Ir2O_{6.5}}                                                & 0.327-0.329   \\
			\ce{Pb2Ir2O6}                                                  & 0.327         \\
			\multicolumn{2}{l}{\ce{Pb2Ir2O7} with \% cation antisites, \hfill average \textit{x}} \\
			\multicolumn{1}{c}{12.25\%}                                   & 0.336             \\
			\multicolumn{1}{c}{25.0\%}                                   & 0.337             \\
			\multicolumn{1}{c}{37.5\%}                                   & 0.344             \\ 
			\multicolumn{1}{c}{50.0\%}                                   & 0.342             \\ \hline
		\end{tabular}
		\caption{Oxygen position parameter \textit{x}, describing the oxygen column delocalization in the pyrochlore structure, calculated from DFT structural optimizations with strain, interface, and compositional variations.}
		\label{tab:dft}
	\end{table}
	\subsection{Conclusion}
	This paper has demonstrated a new electron microscopy technique capable of imaging light and heavy atoms in thicker, strongly scattering electron microscopy specimens using 4D-STEM data recorded in a single specimen orientation. 
	For the case of the \PIO-\ZrO\ heterostructure we explored the visualisation of the \PIO\ and \ZrO\ layers using the $\Smat$-matrix technique. 
	Existing alternatives to $\Smat$-matrix reconstruction and depth sectioning yield similar and supportive results though are not able, on their own, to give definitive evidence of the \PIO\ layer.
	
	\noindent\small\color{Maroon}\textbf{Acknowledgements }\color{Black}
	Work at the Molecular Foundry was supported by the Office of Science, Office of Basic Energy Sciences, of
	the U.S. Department of Energy under Contract No. DE-AC02-05CH11231. C.O. acknowledges support from the U.S. Department of Energy Early Career Research Program. J.C. and H.G.B. acknowledge support from the Presidential Early Career Award for Scientists and Engineers (PECASE) through the U.S. Department of Energy. S.M.G., E.S. and K.I. also acknowledge support from the Laboratory Directed Research and Development Program of LBNL under the U.S. Department of Energy Contract No. DE-AC02-05CH11231. E.S. acknowledges support from the US-Irish Fulbright Commission and work supported by the Air Force Office of Scientific Research under award number FA9550-18-1-0480. Computational resources were provided by the National Energy Research Scientific Computing Center and the Molecular Foundry. This research was partly supported under the Discovery Projects funding scheme of the Australian Research Council (Project No. FT190100619).
	%
	M.S., T.C.P. and C.T.K acknowledge support from the Deutsche Forschungsgemeinschaft (DFG, German Research Foundation) - Project-ID 414984028 - SFB 1404.
	
	All software used in this publication is available in the supplementary material. Due to their large size experimental datasets are not hosted along with the publication but can readily be made available upon request.
	\normalsize
	%
	

\end{document}